\documentclass[aps]{revtex4}  
\usepackage{graphicx}  
\usepackage{dcolumn}   
\usepackage{bm}        
\usepackage{amssymb}   
\def\MET{{\mbox{$E\kern-0.57em\raise0.19ex\hbox{/}_{T}$}}}
\def\met{{\mbox{$E\kern-0.57em\raise0.19ex\hbox{/}_{T}$}}}
\def\DZ{D\O\ }
\def\DZero{D\O\ }
\def\Dzero{D\O\ }

\def\ifb{~fb$^{-1}$}
\def\pp{$p\bar{p}$}

\def\ttbar{$t\bar{t}$}

\def\lmet{$WH\rightarrow \ell\kern-0.45em\raise0.19ex\hbox{/} \nu b\bar{b}$}

\def\ZHll{$ZH\rightarrow \ell^+ \ell^- b\bar{b}$}

\def\pwww{$p\bar{p}\rightarrow WH \rightarrow WW^{+} W^{-}$}
\def\www{$WH \rightarrow WW^{+} W^{-}$}
\def\phww{$p\bar{p}\rightarrow H \rightarrow W^{+} W^{-}$}
\def\hww{$H\rightarrow W^+ W^-$}
\def\hbb{$H\rightarrow b\bar{b}$}

\def\tevE{$\sqrt{s}=1.96$~TeV}

\begin{document}
\rightline{FERMILAB-PUB-08-270-E}
\rightline{CDF Note 9465}
\rightline{\DZ Note 5754}
\vskip0.5in

\title{Combined CDF and \DZ Upper Limits on 
Standard Model Higgs Boson Production \\at High Mass (155--200 GeV/c$^2$) with  3 fb$^{-1}$ of data\\[2.5cm]}

\author{
The TEVNPH Working Group\footnote{The Tevatron
New-Phenomena and Higgs working group can be contacted at
TEVNPHWG@fnal.gov. More information can be found at http://tevnphwg.fnal.gov/.}
 }
\affiliation{\vskip0.3cm for the CDF and \DZ Collaborations\\
\vskip0.2cm 
\today} 
\begin{abstract}
\vskip0.3in
We combine results from CDF and D\O\ searches for a standard model Higgs
boson ($H$) in  \pp~collisions at the Fermilab Tevatron at
$\sqrt{s}=1.96$~TeV.
 With 3.0\ifb~ of data analyzed at CDF, and
3.0 \ifb~ at D\O, the 95\% C.L. upper limits 
on  Higgs boson production 
are a factor
of 1.2, 1.0 and 1.3  higher than the SM cross section for a Higgs boson mass of
$m_{H}=$165, 170 and 175~GeV/c$^2$, respectively.
We exclude at 95\% C.L. a standard model Higgs boson of $m_H=170$ GeV/$c^2$.
Based on simulation, the ratios of the corresponding median expected upper limit 
to the Standard Model cross section are
1.2,  1.4 and 1.7.  
Compared to the previous Higgs Tevatron combination, more data 
and refined analysis techniques have been used.
These results extend significantly the
individual limits of each experiment and provide new knowledge on the mass of the
standard model Higgs boson beyond the LEP direct searches.\\[2cm]
 {\hspace*{5.5cm}\em Preliminary Results}
\end{abstract}

\maketitle

\newpage
\section{Introduction} 
The search for the last unobserved particle of the standard model (SM),
the Higgs boson, has been a major goal
of High Energy Physics for many years, and
 is  a central part of Fermilab's Tevatron program. 
Direct searches at the CERN LEP collider have set a 95\% C.L. limit on the Higgs boson mass
of $m_H > 114.4$~GeV~\cite{sm-lep}.
Taking into account this limit, precision electroweak measurements  
indirectly constrain the SM Higgs boson mass to 
be lower than 190~GeV  at the 95\% C.L.~\cite{elweak}, which is
within reach of the Fermilab Tevatron collider experiments.

Previous CDF and \Dzero 
results on Higgs searches were combined 
in the Tevatron Higgs combination presented in April 2008~\cite{tev-comb-2008}.
Both CDF and \Dzero 
have recently 
reported new and updated searches for the SM Higgs boson~\cite{cdfWH}-\cite{dzHgg} 
and their combination~\cite{CDFhiggs,DZhiggs}. 

In this note, we combine the most recent 
results of all such searches 
in \pp~collisions at~\tevE\ which are sensitive to a high mass (155-200 GeV/c$^2$) Higgs: 
the searches for a SM Higgs boson decaying to $WW$ pairs (the $W$'s then                     
decaying leptonically) and produced through gluon-gluon fusion (\phww),                      
vector boson fusion (VBF), or in association with vector bosons (\pwww and                   
$p \bar{p} \rightarrow WH/ZH$ with hadronic  $W/Z$ decays)
in data
corresponding to integrated luminosities of 3.0\ifb~at
CDF and 3.0\ifb~at D\O . 

To simplify their combination, 
the searches are separated into 
mutually exclusive 
final states
(see Table~\ref{tab:cdfacc} and ~\ref{tab:dzacc}) 
referred to as ``analyses'' in this note. 
Selection procedures for each
analysis are detailed in Refs.~\cite{cdfHWW,cdfWWW,dzHWW,dzWWW}, and 
are briefly described below.

\section{Acceptance, Backgrounds and Luminosity}  

Event selections are similar for the corresponding CDF and D\O\ analyses.
 For the \hww~analyses, a large \met~and two opposite-signed, isolated leptons
(any combination of electrons or muons) are selected, defining three final states
($e^+e^-$, $e^\pm \mu^\mp$, and $\mu^+\mu^-$) for D\O .
CDF separates the \hww\ events into five non-overlapping samples, first by
separating the events by jet multiplicity (0, 1  or 2), then subdviding the 
0 and 1 jet samples in two, one having a low signal/bacgkround (S/B)
ratio, the other having a higher one.  The presence of
neutrinos in the final state prevents  reconstruction of the
Higgs boson mass, so other variables have to be used for separating 
signal from background. In these analyses, 
the final discriminants are neural-network outputs
based on several kinematic variables~\cite{cdfHWW,dzHWW}. These 
include likelihoods constructed from matrix-element probabilities as input
to the neural network, for  CDF. All analyses in these channels have
been updated with more data and analysis improvements compared to our previous
combination~\cite{tev-comb-2008}.

The CDF and D\O\ experiments 
also contribute  \www~analyses, where the associated $W$ boson and
the $W$ boson from the Higgs boson decay which has the same charge are required 
to decay leptonically,
thereby defining  like-sign dilepton final states 
($e^\pm e^\pm$, $e^\pm \mu^\pm$, and $\mu^{\pm}\mu^{\pm}$)
containing all decays of the third
$W$ boson. In  these analyses, CDF derive the limits from a counting experiment,
while for D\O\
the final variable is a likelihood discriminant formed from several
topological variables.

Higgs boson signals (gluon-gluon fusion, vector boson production, or
associated production with vector bosons) are simulated using \textsc{PYTHIA}
~\cite{pythia}, and \textsc{CTEQ6L}~\cite{cteq}  
parton distribution functions at 
leading-order (LO).
The signal cross sections are
normalized to next-to-next-to-leading order (NNLO)
calculations~\cite{nnlo1,nnlo2}, and branching ratios from
\textsc{HDECAY}~\cite{hdecay}. 
The $gg \rightarrow H$ production cross section is also corrected for two-loop
electroweak corrections~\cite{degrassi}.

For both CDF and D\O , events from
multijet (instrumental) backgrounds (``QCD production'') are measured
in data with different methods, in orthogonal samples.
For CDF, backgrounds
from other SM processes were generated using \textsc{PYTHIA},
\textsc{ALPGEN}~\cite{alpgen}, \textsc{MC@NLO}~\cite{MC@NLO}
 and \textsc{HERWIG}~\cite{herwig}
programs. For D\O , these backgrounds were generated using
\textsc{PYTHIA}, \textsc{ALPGEN}, and \textsc{COMPHEP}~\cite{comphep},
with \textsc{PYTHIA} providing parton-showering and hadronization for
all the generators.  Background processes were normalized using either
experimental data or next-to-leading order calculations from
\textsc{MCFM}~\cite{mcfm}.

Integrated luminosities, 
and references to the Collaborations' public documentation for each analysis
are given in Table~\ref{tab:cdfacc}
for CDF and in Table~\ref{tab:dzacc} for D\O .  
The tables include the ranges of Higgs boson mass ($m_H$) over which
the searches were performed, but the combination presented in this note
is performed only for search of Higgs bosons with mass of 155 GeV/c$^2$ or above.

\begin{table}[h]
\caption{\label{tab:cdfacc}Luminosity, explored mass range and references 
for the CDF analyses.  $\ell$ stands for either $e$ or $\mu$.
}
\begin{ruledtabular}
\begin{tabular}{lcc}
&
  $H\rightarrow W^+ W^- $ &  $WH \rightarrow WW^+ W^-$~~~~~~~~\\ 
 &0,1 jet (low,high S/B), 2jet &  $\rightarrow \ell^\pm\nu \ell^\pm\nu$ \\ \hline 
Luminosity (\ifb)         & 3.0  & 1.9 \\ 
$m_{H}$ range (GeV/c$^2$) &  110-200 &  110-200\\
Reference       &  \cite{cdfHWW} &  \cite{cdfWWW} \\
\end{tabular}
\end{ruledtabular}
\end{table}
\vglue 0.5cm 
\begin{table}[h]
\caption{\label{tab:dzacc}Luminosity, explored mass range and references 
for the D\O\ analyses.  $\ell$ stands for either $e$ or $\mu$.
}
\begin{ruledtabular}
\begin{tabular}{lcc}
 &  $H\rightarrow W^+ W^- $ &  $WH \rightarrow WW^+ W^-$  \\ 
 & $\rightarrow \ell^\pm\nu \ell^\mp\nu$ & 
$\rightarrow \ell^\pm\nu \ell^\pm\nu$   \\ \hline 
Luminosity (\ifb)         &  3.0 & 1.1 \\ 
$m_{H}$ range (GeV/c$^2$) &  110-200 & 120-200 \\
Reference      & \cite{dzHWW}  & \cite{dzWWW}  \\
\end{tabular}
\end{ruledtabular}
\end{table}

\section{Combining Channels} 

To verify that the final result does not depend on the
details of the statistical formulation, 
we performed two types of combinations, using the
Bayesian and  Modified Frequentist approaches, which give similar results
(within 10\% ).
Both
methods rely on distributions in the final discriminants, and not just on
their single integrated
values.  Systematic uncertainties enter as uncertainties on the
expected number of signal and background events, as well
as on the distribution of the discriminants in 
each analysis (``shape uncertainties'').
Both methods use likelihood calculations based on Poisson
probabilities.

\subsection{Bayesian Method}

Because there is no experimental information on the production cross section for
the Higgs boson, in the Bayesian technique~\cite{CDFhiggs} we assign a flat prior
to the total selected Higgs boson cross section.  For a given Higgs boson mass, the
combined likelihood is a product of likelihoods for the individual
channels, each of which is a product over histogram bins:

\begin{equation}
{\cal{L}}(R,{\vec{s}},{\vec{b}}|{\vec{n}},{\vec{\theta}})\times\pi({\vec{\theta}})
= \prod_{i=1}^{N_C}\prod_{j=1}^{Nbins} \mu_{ij}^{n_{ij}} e^{-\mu_{ij}}/n_{ij}!
\times\prod_{k=1}^{n_{np}}e^{-\theta_k^2/2}
\end{equation}

\noindent where the first product is over the number of channels
($N_C$), and the second product is over histogram bins containing
$n_{ij}$ events, binned in  ranges of the final discriminants used for
individual analyses, such as the dijet mass, neural-network outputs, 
or matrix-element likelihoods.
 The parameters that contribute to the
expected bin contents are $\mu_{ij} =R \times s_{ij}({\vec{\theta}}) + b_{ij}({\vec{\theta}})$ 
for the
channel $i$ and the histogram bin $j$, where $s_{ij}$ and $b_{ij}$ 
represent the expected background and signal in the bin, and $R$ is a scaling factor
applied to the signal to test the sensitivity level of the experiment.  
Truncated Gaussian priors are used for each of the nuisance parameters
$\theta_k$, which define
the
sensitivity of the predicted signal and background estimates to systematic uncertainties.
These
can take the form of uncertainties on overall rates, as well as the shapes of the distributions
used for combination.   These systematic uncertainties can be far larger
than the expected SM signal, and are therefore important in the calculation of limits. 
The truncation
is applied so that no prediction of any signal or background in any bin is negative.
The posterior density function is
then integrated over all parameters (including correlations) except for $R$,
and a 95\% confidence level upper limit on $R$ is estimated
by calculating the value of $R$ that corresponds to 95\% of the area
of the resulting distribution.

\subsection{Modified Frequentist Method}

The Modified Frequentist technique relies on the $CL_s$ method, using a
log-likelihood ratio (LLR) as test statistic:
\begin{equation}
LLR = -2\ln\frac{p({\mathrm{data}}|H_1)}{p({\mathrm{data}}|H_0)},
\end{equation}
where $H_1$ denotes the test hypothesis, which admits the presence of 
backgrounds and a Higgs
boson signal, while $H_0$ is the null hypothesis, for
only  backgrounds.  The probabilities $p$ are computed using the best-fit
values of the nuisance parameters for each event, separately 
for each of the two hypotheses,
and include the Poisson probabilities of observing the data multiplied by Gaussian
constraints for the values of the nuisance parameters~\cite{pflh}.  This technique
extends the LEP procedure~\cite{pdgstats} which does not involve a fit, in order to
yield better sensitivity when expected signals are small and
systematic uncertainties on backgrounds are large. 

The $CL_s$ technique involves computing two $p$-values, $CL_{s+b}$ and $CL_b$.
The latter is defined by
\begin{equation}
1-CL_b = p(LLR\le LLR_{\mathrm{obs}} | H_0),
\end{equation}
where $LLR_{\mathrm{obs}}$ is the value of the test statistic computed for the
data. $1-CL_b$ is the probability of observing a signal-plus-background-like outcome 
without the presence of signal, i.e. the probability
that an upward fluctuation of the background provides  a signal-plus-background-like
response as observed in data.
The other $p$-value is defined by
\begin{equation}
CL_{s+b} = p(LLR\ge LLR_{\mathrm{obs}} | H_1),
\end{equation}
and this corresponds to the probability of a downward fluctuation of the sum
of signal and background in 
the data.  A small value of $CL_{s+b}$ reflects inconsistency with  $H_1$.
It is also possible to have a downward fluctuation in data even in the absence of
any signal, and a small value of $CL_{s+b}$ is possible even if the expected signal is
so small that it cannot be tested with the experiment.  To minimize the possibility
of  excluding  a signal to which there is insufficient sensitivity 
(an outcome  expected 5\% of the time at the 95\% C.L., for full coverage),
we use the quantity $CL_s=CL_{s+b}/CL_b$.  If $CL_s<0.05$ for a particular choice
of $H_1$, that hypothesis is deemed excluded at the 95\% C.L.

Systematic uncertainties are included  by fluctuating the predictions for
signal and background rates in each bin of each histogram in a correlated way when
generating the pseudoexperiments used to compute $CL_{s+b}$ and $CL_b$.

\subsection{Systematic Uncertainties} 

Systematic uncertainties differ
between experiments and analyses, and they affect the rates and shapes of the predicted
signal and background in correlated ways.  The combined results incorporate
the sensitivity of predictions to  values of nuisance parameters,
and correlations are included, between rates and shapes, between signals and backgrounds,
and between channels within experiments and between experiments.
More on these issues can be found in the
individual analysis notes~\cite{cdfWH}-\cite{dzHgg}.  Here we
consider only the largest contributions and correlations between and
within the two experiments.

\subsubsection{Correlated Systematics between CDF and D\O}
The uncertainty on the measurement of the integrated luminosity is 6\%
(CDF) and 6.1\% (D\O ).  
Of this value, 4\% arises from the uncertainty
on the inelastic \pp~scattering cross section, which is correlated
between CDF and D\O . 
The uncertainty on the production rates for the signal, for 
top-quark processes (\ttbar~and single top) and for electroweak processes
($WW$, $WZ$, and $ZZ$) are taken as correlated between the two
experiments. As the methods of measuring the multijet (``QCD'')
backgrounds differ between CDF and D\O , there is no
correlation assumed between these rates.  
The calibrations of fake leptons and unvetoed $\gamma\rightarrow e^+e^-$ conversions,
are performed by each collaboration
using independent data samples and methods, hence are considered uncorrelated.

\subsubsection{Correlated Systematic Uncertainties for CDF}
The dominant systematic uncertainties for the CDF analyses are shown in
Tables~\ref{tab:cdfsystww0}, \ref{tab:cdfsystww1},  \ref{tab:cdfsystww2}, 
and \ref{tab:cdfsystwww}.
Each source of uncertainty induces a correlated uncertainty across all CDF channels
sensitive to that source.
For
\hww, the largest uncertainty comes from
MC modeling (5\%).  For simulated backgrounds, the uncertainties on the
expected rates range from 11-40\% (depending on background). 
The backgrounds with the largest systematic uncertainties                                          
are in general quite small. Such uncertainties are 
constrained by fits to  the nuisance parameters, and they
do not affect the result significantly.
Because
the \hww~channel, the  uncertainty on luminosity is taken to be correlated
between signal and background. The differences in the resulting limits,
whether treating the
remaining uncertainties as correlated or uncorrelated  is within $5\%$.

\subsubsection{Correlated Systematic Uncertainties for D\O }
The dominant systematic uncertainties for D\O\ analyses are shown in Tables
\ref{tab:d0systww} and \ref{tab:d0systwww}.
Each source of uncertainty induces a correlated uncertainty across all D\O\ channels
sensitive to that source.
For  \hww~and \www, the largest uncertainties are associated with lepton
measurement and acceptance. These values range from 2-11\% depending on
the final state.  The largest contributing factor to all analyses is
the uncertainty on cross sections for simulated background, and is 
6-18\%. 
All systematic uncertainties arising from the
same source are taken to be correlated between the different backgrounds and
between signal and background.


\begin{table}
\begin{center}
\caption{Systematic uncertainties on the contributions for CDF's
$H\rightarrow W^+W^-\rightarrow\ell^{\pm}\ell^{\prime \mp}$ channel
with zero jets.  Systematic uncertainties are listed by name, see 
the original references for a detailed explanation of their meaning 
and on how they are derived.  
Systematic uncertainties for $gg \to H$ shown in this table are obtained for $m_H=160$ GeV/c$^2$.
Uncertainties are relative, in percent and are symmetric unless otherwise indicated.
Systematic in bold are correlated across jet bins but                   
not across channels.  
Systematics in italics are correlated across jet bins and                  
across appropriate channels.
}
\label{tab:cdfsystww0}
\begin{footnotesize}
\vskip 0.8cm                                                                                                         
{\centerline{CDF:   $H \rightarrow WW \rightarrow \ell^{\pm} \ell^{\prime \mp}$ + 0 Jets Analysis}}
\vskip 0.099cm                                                                                                         
\begin{tabular}{|l|c|c|c|c|c|c|c|c|c|c|c|} \hline
Uncertainty Source         &  $WW$      &  $WZ$         &  $ZZ$  &  $t\bar{t}$  &  DY           &  $W\gamma$   & $W$+jet &$gg\to H$&  $WH$  &  $ZH$  &  VBF  \\ \hline 
{\bf Cross Section}        &            &               &               &        &              &        &         &        &        &        &         \\ \hline
Scale                      &            &               &               &        &              &        &         & 10.9\% &        &        &         \\ 
PDF Model                  &            &               &               &        &              &        &         &  5.1\% &        &        &         \\ 
Total                      & {\it 10.0\%} & {\it 10.0\%} & {\it 10.0\%} & 15.0\% &  5.0\%       & 10.0\% &         & 12.0\% &        &        &         \\ 
{\bf Acceptance}           &            &               &               &        &              &        &         &        &        &        &         \\ \hline
Scale (leptons)            &            &               &               &        &              &        &         &  2.5\% &        &        &         \\ 
Scale (jets)               &            &               &               &        &              &        &         &  4.6\% &        &        &         \\ 
PDF Model (leptons)        &  1.9\%     &  2.7\%        &  2.7\%        &  2.1\% &  4.1\%       &  2.2\% &         &  1.5\% &        &        &         \\ 
PDF Model (jets)           &            &               &               &        &              &        &         &  0.9\% &        &        &         \\ 
Higher-order Diagrams      & {\bf 5.5\%} & {\bf 10.0\%} & {\bf 10.0\%}  & 10.0\% & {\bf 5.0\%}  & {\bf 10.0\%} &   &        &        &        &         \\ 
Missing Et Modeling        &  1.0\%     &  1.0\%        &  1.0\%        &  1.0\% & 20.0\%       &  1.0\% &         &  1.0\% &        &        &         \\ 
Conversion Modeling        &            &               &               &        &              & 20.0\% &         &        &        &        &         \\ 
Jet Fake Rates             &            &               &               &        &              &        &         &        &        &        &         \\ 
(Low S/B)                  &            &               &               &        &              &        & 21.5\%  &        &        &        &         \\
(High S/B)                 &            &               &               &        &              &        & 27.7\%  &        &        &        &         \\ 
MC Run Dependence          &  3.9\%     &               &               &  4.5\% &              &  4.5\% &         &  3.7\% &        &        &         \\ 
Lepton ID Efficiencies     &  2.0\%     &  1.7\%        &  2.0\%        &  2.0\% &  1.9\%       &  1.4\% &         &  1.9\% &        &        &         \\ 
Trigger Efficiencies       &  2.1\%     &  2.1\%        &  2.1\%        &  2.0\% &  3.4\%       &  7.0\% &         &  3.3\% &        &        &         \\ 
{\bf Luminosity}           &  5.9\%     &  5.9\%        &  5.9\%        &  5.9\% &  5.9\%       &  5.9\% &         &  5.9\% &        &        &         \\ \hline
\end{tabular}
\end{footnotesize}
\end{center}
\end{table}

\begin{table}
\begin{center}
\caption{Systematic uncertainties on the contributions for CDF's
$H\rightarrow W^+W^-\rightarrow\ell^{\pm}\ell^{\prime \mp}$ channel
with one jet.  Systematic uncertainties are listed by name, see 
the original references for a detailed explanation of their meaning 
and on how they are derived.  
Systematic uncertainties for $gg \to H$ shown in this table are obtained for $m_H=160$ GeV/c$^2$.
Uncertainties are relative, in percent and are symmetric unless otherwise indicated.
Negative numbers indicate systematics that are anti-correlated between                       
channels.  
Systematic in bold are correlated across jet bins but                   
not across channels.  
Systematics in italics are correlated across jet bins and                  
across appropriate channels.
}
\label{tab:cdfsystww1}
\begin{footnotesize}
\vskip 0.8cm                                                                                                         
{\centerline{CDF:   $H \rightarrow WW \rightarrow \ell^{\pm} \ell^{\prime \mp}$ + 1 Jet Analysis}}
\vskip 0.099cm                                                                                                         
\begin{tabular}{|l|c|c|c|c|c|c|c|c|c|c|c|} \hline
Uncertainty Source         &  $WW$      &  $WZ$         &  $ZZ$  &  $t\bar{t}$   &  DY          & $W\gamma$   & $W$+jet &$gg \to H$&  $WH$  &  $ZH$  &  VBF \\ \hline 
{\bf Cross Section}        &            &               &               &        &              &        &         &        &        &        &         \\ \hline
Scale                      &            &               &               &        &              &        &              & 10.9\% &        &        &        \\ 
PDF Model                  &            &               &               &        &              &        &              &  5.1\% &        &        &        \\ 
Total                      & {\it 10.0\%} & {\it 10.0\%} & {\it 10.0\%} & 15.0\% &  5.0\%       & 10.0\% &              & 12.0\% &  5.0\% &  5.0\% & 10.0\% \\ 
{\bf Acceptance}           &            &               &               &        &              &        &         &        &        &        &         \\ \hline
Scale (leptons)            &            &               &               &        &              &        &              &  2.8\% &        &        &        \\ 
Scale (jets)               &            &               &               &        &              &        &              & -5.1\% &        &        &        \\ 
PDF Model (leptons)        &  1.9\%     &  2.7\%        &  2.7\%        &  2.1\% &  4.1\%       &  2.2\% &              &  1.7\% &  1.2\% &  0.9\% &  2.2\% \\ 
PDF Model (jets)           &            &               &               &        &              &        &              & -1.9\% &        &        &        \\ 
Higher-order Diagrams      &  {\bf 5.5\%} & {\bf 10.0\%} & {\bf 10.0\%} & 10.0\% & {\bf 5.0\%}  & {\bf 10.0\%} &        &        & {\it 10.0\%} & {\it 10.0\%} & {\it 10.0\%} \\ 
Missing Et Modeling        &  1.0\%     &  1.0\%        &  1.0\%        &  1.0\% & 20.0\%       &  1.0\% &              &  1.0\% &  1.0\% &  1.0\% &  1.0\% \\ 
Conversion Modeling        &            &               &               &        &              & 20.0\% &              &        &        &        &        \\ 
Jet Fake Rates             &            &               &               &        &              &        &              &        &        &        &        \\ 
(Low S/B)                  &            &               &               &        &              &        & 22.2\%       &        &        &        &        \\
(High S/B)                 &            &               &               &        &              &        & 31.5\%       &        &        &        &        \\ 
MC Run Dependence          &  1.8\%     &               &               &  2.2\% &              &  2.2\% &              &  2.6\% &  2.6\% &  1.9\% &  2.8\% \\ 
Lepton ID Efficiencies     &  2.0\%     &  2.0\%        &  2.2\%        &  1.8\% &  2.0\%       &  2.0\% &              &  1.9\% &  1.9\% &  1.9\% &  1.9\% \\ 
Trigger Efficiencies       &  2.1\%     &  2.1\%        &  2.1\%        &  2.0\% &  3.4\%       &  7.0\% &              &  3.3\% &  2.1\% &  2.1\% &  3.3\% \\ 
{\bf Luminosity}           &  5.9\%     &  5.9\%        &  5.9\%        &  5.9\% &  5.9\%       &  5.9\% &              &  5.9\% &  5.9\% &  5.9\% &  5.9\% \\ \hline
\end{tabular}
\end{footnotesize}
\end{center}
\end{table}

\begin{table}
\begin{center}
\caption{Systematic uncertainties on the contributions for CDF's
$H\rightarrow W^+W^-\rightarrow\ell^{\pm}\ell^{\prime \mp}$ channel
with two or more jets.  Systematic uncertainties are listed by name, see 
the original references for a detailed explanation of their meaning 
and on how they are derived.  
Systematic uncertainties for $gg \to H$ shown in this table are obtained for $m_H=160$ GeV/c$^2$.
Uncertainties are relative, in percent and are symmetric unless otherwise indicated.
Systematic in bold are correlated across jet bins but                   
not across channels.  
Systematics in italics are correlated across jet bins and                  
across appropriate channels.
}
\label{tab:cdfsystww2}
\begin{footnotesize}
\vskip 0.8cm                                                                                                         
{\centerline{CDF:   $H \rightarrow WW \rightarrow \ell^{\pm} \ell^{\prime \mp}$ + $\ge2$ Jets Analysis}}
\vskip 0.099cm                                                                                                         
\begin{tabular}{|l|c|c|c|c|c|c|c|c|c|c|c|} \hline
Uncertainty Source         &  $WW$      &  $WZ$         &  $ZZ$  &  $t\bar{t}$  &  DY           &  $W\gamma$    & $W$+jet &$gg\to H$&  $WH$  &  $ZH$  &  VBF    \\ \hline 
{\bf Cross Section}        &            &               &               &        &              &        &         &        &        &        &         \\ \hline
Scale                      &            &               &               &        &              &               &         & 10.9\% &        &        &          \\ 
PDF Model                  &            &               &               &        &              &               &         &  5.1\% &        &        &          \\ 
Total                      & {\it 10.0\%} & {\it 10.0\%} & {\it 10.0\%} & 15.0\% &  5.0\%       & 10.0\%        &         & 12.0\% &  5.0\% &  5.0\% & 10.0\%       \\ 
{\bf Acceptance}           &            &               &               &        &              &        &         &        &        &        &         \\ \hline
Scale (leptons)            &            &               &               &        &              &               &         &  3.1\% &        &        &          \\ 
Scale (jets)               &            &               &               &        &              &               &         & -8.7\% &        &        &          \\ 
PDF Model (leptons)        &  1.9\%     &  2.7\%        &  2.7\%        &  2.1\% &  4.1\%       &  2.2\%        &         &  2.0\% &  1.2\% &  0.9\% &  2.2\%   \\ 
PDF Model (jets)           &            &               &               &        &              &               &         & -2.8\% &        &        &          \\ 
Higher-order Diagrams      & {\bf 10.0\%} & {\bf 10.0\%} & {\bf 10.0\%} & 10.0\% & {\bf 10.0\%} & {\bf 10.0\%}  &         &        & {\it 10.0\%} & {\it 10.0\%} & {\it 10.0\%}\\
Missing Et Modeling        &  1.0\%     &  1.0\%        &  1.0\%        &  1.0\% & 20.0\%       &  1.0\%        &         &  1.0\% &  1.0\% &  1.0\% &  1.0\%   \\ 
Conversion Modeling        &            &               &               &        &              & 20.0\%        &         &        &        &        &          \\ 
$b$-tag Veto               &            &               &               &  7.0\% &              &               &         &        &        &        &          \\ 
Jet Fake Rates             &            &               &               &        &              &               & 27.1\%  &        &        &        &          \\ 
MC Run Dependence          &  1.0\%     &               &               &  1.0\% &              &  1.0\%        &         &  1.7\% &  2.0\% &  1.9\% &  2.6\%   \\ 
Lepton ID Efficiencies     &  1.9\%     &  2.9\%        &  1.9\%        &  1.9\% &  1.9\%       &  1.9\%        &         &  1.9\% &  1.9\% &  1.9\% &  1.9\%   \\ 
Trigger Efficiencies       &  2.1\%     &  2.1\%        &  2.1\%        &  2.0\% &  3.4\%       &  7.0\%        &         &  3.3\% &  2.1\% &  2.1\% &  3.3\%   \\ 
{\bf Luminosity}           &  5.9\%     &  5.9\%        &  5.9\%        &  5.9\% &  5.9\%       &  5.9\%        &         &  5.9\% &  5.9\% &  5.9\% &  5.9\%   \\ \hline
\end{tabular}                                                                                                             
\end{footnotesize}
\end{center}
\end{table}

\begin{table}
\begin{center}
\caption{Systematic uncertainties on the contributions for D\O 's
$H\rightarrow WW \rightarrow\ell^{\pm}\ell^{\prime \mp}$ channel.
Systematic uncertainties are listed by name, see the original references for a detailed explanation of their meaning and on how they are 
derived Systematic uncertainties shown in this table are obtained for the $m_H=160$ GeV/c$^2$ Higgs selection.
Uncertainties are relative, in percent.}
\label{tab:d0systww}
\vskip 0.8cm                                                                                                          
{\centerline{D\O : $H\rightarrow WW \rightarrow\ell^{\pm}\ell^{\prime \mp}$ Analysis }}
 \vskip 0.099cm      
\begin{tabular}{| l | c | c | c | c | c | c | c|c|c|c|}                                                                               
\hline 
& $\Sigma$ Bkgd  & Signal & $Z\rightarrow \ell\ell$ & $Z\rightarrow \tau\tau$  & $W+jets/\gamma$ &~~~~$t\bar{t}$~~~~ &~~~~ZZ~~~~ &~~~~WZ~~~~ &~~~~WW~~~~ & QCD\\\hline
Jet energy scale        &    7    &    3   &    7    &   6     &  6     &  7   &    2   &   1   &    1  &    - \\
Jet energy resol.       &    3    &    3   &    3    &   4     &  4     &  3   &    3   &   3   &    2  &    - \\
Jet ID                  &  4      &  0     &    5   &   13     &  6     &  2   &    1   &  4    &    0  &    - \\
${\cal L}$ Reweighting  &  0      &  3     &    0   &    13    &  3     &  1   &    0   &  0    &    0  &    - \\
Beam Reweighting        &  2      &  1     &    2   &    6     &  4     &  1   &    1   &  2    &    1  &    - \\
$Z-p_T$ Rew             &  13     &   0    &   17   &  14      &  0     &  0   &    0   &  0    &    0  &    - \\
Leptom momentum         &   4     &   0    &   5    &  4       &  3     &  2   &    1   &   2   &    0  &    - \\
Cross section           &  -      &     10 &      6 &        6 &     20 &   10 &      6 &      6&    6  &     - \\
$QCD$                   &  -      &      - &      - &        - &      - &    - &      - &     - &     - &     30 \\
Normalization           &  7      &      7 &      - &        - &      - &    - &      - &     - &     - &     - \\
Lepton ID               &  4      &      4 &      - &        - &      - &    - &      - &     - &     - &     - \\
\hline                                                                                                                                   
\end{tabular}                                                                                                           
\end{center}
\end{table}


\begin{table}
\begin{center}
\caption{Systematic uncertainties on the contributions for CDF 's
$WH \rightarrow WWW \rightarrow\ell^{\prime \pm}\ell^{\prime \pm}$ channel.
Systematic uncertainties are listed by name, see the original references for a detailed
 explanation of their meaning and on how they are derived. 
Systematic uncertainties for $WH$ shown in this table are obtained for $m_H=160$ GeV/c$^2$.
The diboson
contribution to the final selected sample is negligible.
Uncertainties are relative, in percent.}
\label{tab:cdfsystwww}
\vskip 0.8cm 
{\centerline{CDF : $WH \rightarrow WWW \rightarrow\ell^{\pm}\ell^{\prime\pm}$ Analysis.}}
\vskip 0.099cm                    
   \begin{tabular}{|l|c|c|c| }
\hline
Contribution & $\gamma$ Conversions & Fakes &  WH \\ \hline
Statistaical Uncertainty &   0      &  0    &  1.6 \\
Fake Rate                &   0      &  25   &  0  \\
Conversions              &  25      &  0    &  0  \\
Luminosity               &  0       &  0    &   6  \\
ISR                      &  0       &  0    &  4.0 \\
FSR                      &  0       &  0    &  4.1 \\
PDF                      &  0       &  0    &  1.9  \\ 
Lepton ID                &  0       &  0    &  1.2 \\ \hline
\end{tabular}
\end{center}
\end{table}

\begin{table}
\begin{center}
\caption{Systematic uncertainties on the contributions for D\O 's
$WH \rightarrow WWW \rightarrow\ell^{\prime \pm}\ell^{\prime \pm}$ channel.
Systematic uncertainties are listed by name, see the original references for a detailed explanation of their meaning and on how they are derived. 
Systematic uncertainties for $WH$ shown in this table are obtained for $m_H=160$ GeV/c$^2$.
Uncertainties are relative, in percent and are symmetric unless otherwise indicated.   }
\label{tab:d0systwww}
\vskip 0.8cm                                                                                                          
{\centerline{D\O : $WH \rightarrow WWW \rightarrow\ell^{\pm}\ell^{\prime\pm}$ Analysis.}}
\vskip 0.099cm                                                                                                          
\begin{tabular}{| l | c | c | c | c | }
\hline
Contribution                           & ~~WZ/ZZ~~ & Charge flips & ~~~QCD~~~ &~~~~~WH~~~~~  \\ \hline
Trigger eff.                           &  5    &  0                     &  0  &  5    \\
Lepton ID/Reco. eff                    & 10    &  0                     &  0  & 10    \\
Cross Section                          &  7    &  0                     &  0  &  6    \\
Normalization                          &  6    &  0                     &  0  &  0    \\ 
Instrumental-ee ($ee$ final state)                                                    
                                       &  0    &  32                    &  15 &  0    \\
Instrumental-em ($e\mu$ final state)                                                  
                                       &  0    &  0                     &  18 &  0    \\
Instrumental-mm ($\mu\mu$ final state)                                                 
                                       &  0    &  $^{+290}_{-100}$      & 32  &  0    \\ \hline
\end{tabular}
\end{center}
\end{table}

\clearpage

 \begin{figure}[t]
 \begin{centering}
 \includegraphics[width=14.0cm]{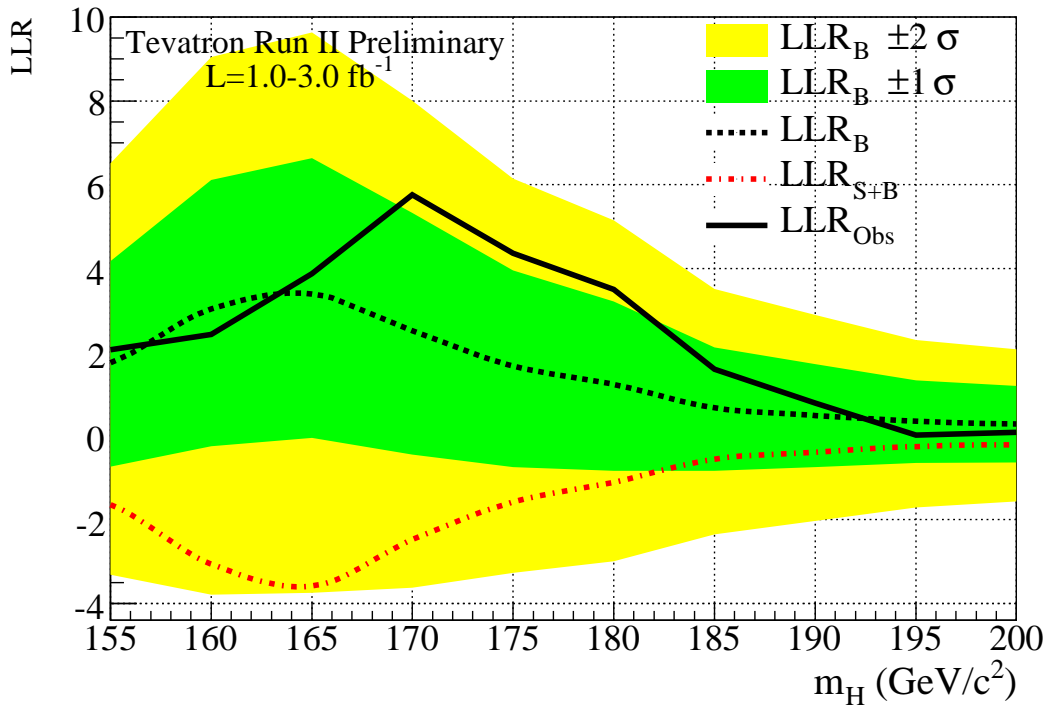}
 \caption{
 \label{fig:comboLLR}
 Distributions of LLR as a function of the Higgs boson mass 
(in steps of 5 GeV/c$^2$)
 for the combination of the
 CDF and D\O\ analyses. }
 \end{centering}
 \end{figure}

\vspace*{1cm}
\section{Combined Results} 

Before extracting the combined limits we study the distributions 
of the 
log-likelihood ratio (LLR) for different hypotheses,
to check the expected
sensitivity across the mass range tested.
Figure~\ref{fig:comboLLR}
displays the LLR distributions
for the combined
analyses as a function of $m_{H}$. Included are the results for the
background-only hypothesis (LLR$_{b}$), the signal and background
hypothesis (LLR$_{s+b}$), and for the data (LLR$_{obs}$).  The
shaded bands represent the 1 and 2 standard deviation ($\sigma$)
departures for LLR$_{b}$. 

These
distributions can be interpreted as follows:
The separation between LLR$_{b}$ and LLR$_{s+b}$ provides a
measure of the discriminating power of the search; 
 the size of the 1- and 2-$\sigma$ LLR$_{b}$ bands
provides an estimate of how sensitive the
analysis is to a signal-plus-background-like fluctuation in data, taking account of
the systematic uncertainties;
the value of LLR$_{obs}$ relative to LLR$_{s+b}$ and LLR$_{b}$
indicates whether the data distribution appears to be more signal-plus-background-like
(i.e. closer to the LLR$_{s+b}$ distribution, which is negative by
construction)
or background-like; the significance of any departures
of LLR$_{obs}$ from LLR$_{b}$ can be evaluated by the width of the
LLR$_{b}$ bands.

Using the combination procedures outlined in Section III, we extract limits on
SM Higgs boson production $\sigma \times B(H\rightarrow X)$ in
\pp~collisions at $\sqrt{s}=1.96$~TeV for $m_H=155-200$ GeV/c$^2$.
To facilitate comparisons with the standard model and to accommodate analyses with
different degrees of sensitivity, we present our results in terms of
the ratio of obtained limits  to  cross section in the SM, as a function of
Higgs boson mass, for test masses for which
both experiments have performed dedicated searches in different channels.
A value of the combined limit ratio which is less or equal to one would indicate that
that particular Higgs boson mass is excluded at the
95\% C.L. 

The combinations of results of each single experiment,
yield the following ratios of 95\% C.L. observed (expected) limits to the SM 
cross section: 
1.6~(1.6) for CDF and 2.0~(1.9) for D\O\ at $m_{H}=165$~GeV/c$^2$, and 
1.8~(1.9) for CDF and 1.7~(2.3) for D\O\ at $m_{H}=170$~GeV/c$^2$.

The ratios of the 95\% C.L. expected and observed  limit to the
SM cross section 
are shown in
Figure~\ref{fig:comboRatio}
for the combined CDF and D\O\ analyses.  The observed and median expected ratios
are listed for the tested Higgs boson masses in Tables~\ref{tab:ratios} and~\ref{tab:ratios-2}, 
with
observed (expected) values of 
1.2 (1.2) at $m_{H}=165$~GeV/c$^2$,
1.0 (1.4) at $m_{H}=170$~GeV/c$^2$, and
1.3 (1.7) at $m_{H}=175$~GeV/c$^2$.
We exclude at the 95\% C.L. the
production of a standard model  Higgs boson with  mass of 170 GeV/c$^2$.
This result is obtained with
both Bayesian and $CL_S$ calculations.
\begin{figure}[hb]
\begin{centering}
\includegraphics[width=16.5cm]{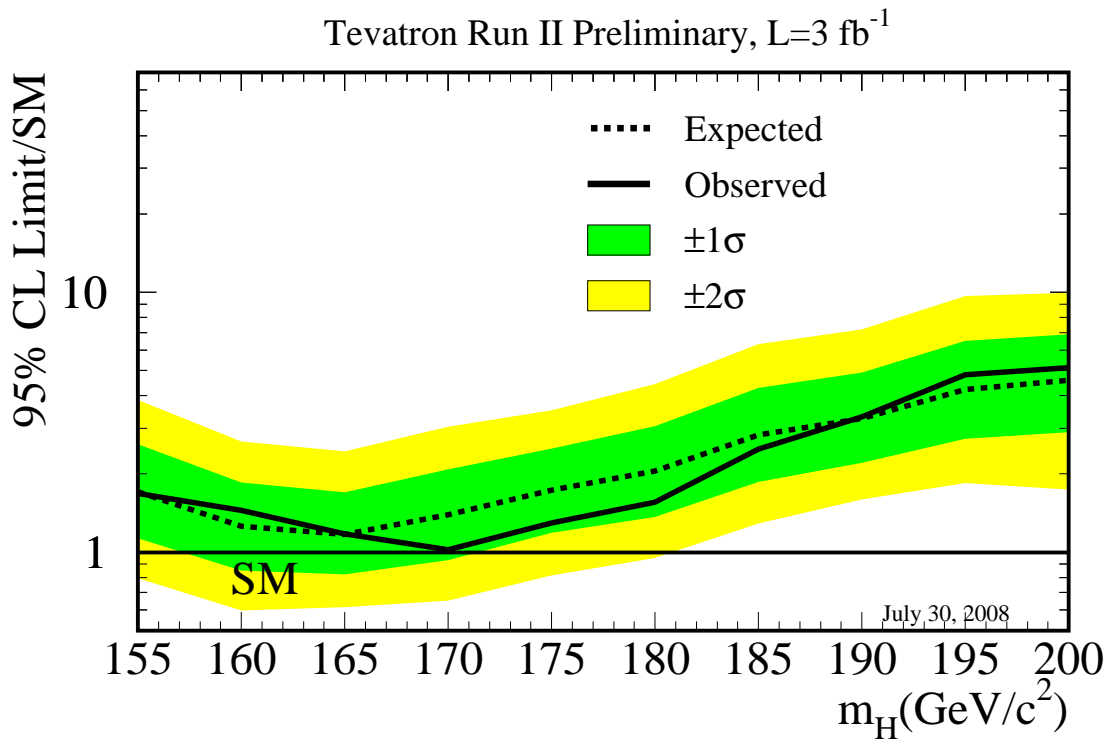}
\caption{
\label{fig:comboRatio}
Observed and expected (median, for the background-only hypothesis)
  95\% C.L. upper limits on the ratios to the
SM cross section, 
as functions of the Higgs boson mass 
for the combined CDF and D\O\ analyses.
The limits are expressed as a multiple of the SM prediction
for test masses (every 5 GeV/$c^2$)
for which both experiments have performed dedicated
searches in different channels.
The points are joined by straight lines 
for better readability.
  The bands indicate the
68\% and 95\% probability regions where the limits can
fluctuate, in the absence of signal. 
The limits displayed in this figure
are obtained with the Bayesian calculation.
}
\end{centering}
\end{figure}

\begin{table}[ht]
\caption{\label{tab:ratios} Ratios of median expected and observed 95\% CL
limit to the SM cross section for the combined CDF and D\O\ analyses as a function
of the Higgs boson mass in GeV/c$^2$, obtained with the Bayesian method.}
\begin{ruledtabular}
\begin{tabular}{lccccccccccc}
                     &    155  & 160 & 165 & 170 & 175 & 180 & 185 & 190 & 195 & 200\\ \hline 
Expected             &   1.7  & 1.3 & 1.2 & 1.4 & 1.7 & 2.0 & 2.8 & 3.3 & 4.2 & 4.6\\
Observed             &   1.7  & 1.4 & 1.2 & 1.0 & 1.3 & 1.6 & 2.5 & 3.3 & 4.8 & 5.1\\
\end{tabular}
\end{ruledtabular}
\end{table}
\begin{table}[ht]
\caption{\label{tab:ratios-2} Ratios of median expected and observed 95\% CL
limit to the SM cross section for the combined CDF and D\O\ analyses as a function
of the Higgs boson mass in GeV/c$^2$, obtained with the $CL_S$ method.}
\begin{ruledtabular}
\begin{tabular}{lcccccccccccc}
               &         155  & 160 & 165 & 170 & 175 & 180 & 185 & 190 & 195 & 200\\ \hline 
Expected       &         1.6  & 1.2 & 1.1 & 1.3 & 1.7 & 2.0 & 2.8 & 3.4 & 4.2 & 4.7\\
Observed       &         1.6  & 1.3 & 1.1 & 0.95& 1.2 & 1.4 & 2.3 & 3.2 & 4.7 & 5.0\\
\end{tabular}
\end{ruledtabular}
\end{table}

We also show in Figure~\ref{fig:comboLLR-2} the 1-$CL_S$ distribution as a function of 
the Higgs boson mass, which is directly interpreted
as the level of exclusion of our search. For instance,  both our observed and expected
results exclude a Higgs boson with $m_H= $ 165 GeV/$c^2$ at $\approx$ 92\% C.L.

 \begin{figure}[t]
 \begin{centering}
 \includegraphics[width=14.0cm]{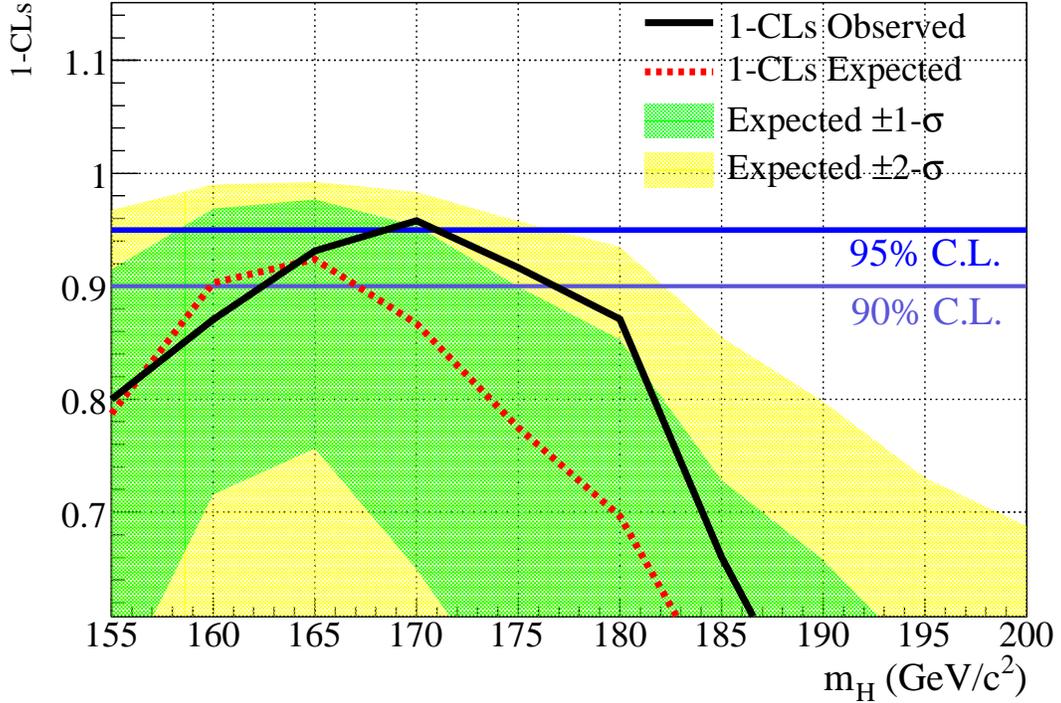}
 \caption{
 \label{fig:comboLLR-2}
 Distributions of 1-$CL_S$ as a function of the Higgs boson mass 
(in steps of 5 GeV/c$^2$)
 for the combination of the
 CDF and D\O\ analyses. }
 \end{centering}
 \end{figure}


\end{document}

Limits observed and expected for CDF search channels (multiples of SM)

HWW

mh    obs    exp
110  151.2  122.6
120  33.9   37.4
130  17.0  17.4
140  9.5   10.7
150  5.7  8.0
160  3.4  4.8
170  3.3  4.9
180  6.8  6.6
190  14.6  9.8
200  18.4  12.9

bb+MET   (WH and ZH signal together)

mh    obs   exp
110   18.5  9.3
115   19.7  9.7
120   22.6  11.5
125   26.6  13.4
130   33.4  16.6
135   43.0  21.0
140   61.5  31.5
150   127.0  72.1

WH->lvbb

mh   obs   exp
110  8.8   8.6
115  10.1  10.0
120  12.0  11.9
130  18.8  17.4
140  40.0  33.0
150  113.0  80.6

ZH->llbb

mh   obs   exp
100  12    13
110  14    14
115  16    16
120  18    18
130  29    28
140  65    54
150  160   140

       105   115     125   135     145
Obs (11.2)  (17.8)  (30.4) (51.4) (103.5)
Exp (14.9)  (20.4)  (27.3) (42.8)  (88.0)

exp  2.8     2.5     2.3   2.0
     2.6     2.7     2.5   2.3                                                                                                            
HWW_v1.7.pdf
mh [GeV]    120  140 160 180 200
exp
combination 28.7 8.3 3.5 5.3 11.7
obs
combination 48.9 12.3 3.1 5.5 11.4

\ $m_H$   \      &  \ expected \ & \ observed \ \\
\ (GeV) \  & \ limit (pb) \ & \ limit (pb)  \  \\ \hline
105   &  1.29   &   1.42 \\
115   &  1.16   &   1.42 \\
125   &  1.12   &   1.41 \\
135   &  0.94   &   1.16 \\
145   &  0.84   &   1.06 \\ \hline

\vglue 0.2cm 
\begin{table}[h]
\caption{\label{tab:dzacc1}The luminosity, mass range explored and reference 
for the D\O\ \hbb~analyses.   $\ell$ stands for $e$ or $\mu$.
}
\begin{ruledtabular}
\begin{tabular}{lcccccccc}
\\
&$WH\rightarrow e\nu b\bar{b}$ & $WH\rightarrow \mu\nu b\bar{b}$  & \lmet  & $ZH\rightarrow \nu\bar{\nu} b\bar{b}$ & $ZH\rightarrow \ell^+\ell^- b\bar{b}$ \\ 
& DT(ST) & DT(ST) &DT(ST) &  DT(ST) & \\\hline
Luminosity (\ifb)         & 1.7 & 1.7 & 0.9 & 0.9& 1.1\\ 
$m_{H}$ range (GeV/c$^2$)       & 105-145 & 105-145 & 105-135 &105-135  & 105-145\\
Reference       & \cite{dzWHl} & \cite{dzWHl}& \cite{dzZHv} & \cite{dzZHv} & \cite{dzZHll} \\
\end{tabular}
\end{ruledtabular}
\end{table}
\vglue 0.2cm 
\begin{table}[h]
 \caption{\label{tab:dzacc3}The luminosity, mass range explored, and reference 
for the D\O\ \www\ and \hww~analyses.  
}
\begin{ruledtabular}
\begin{tabular}{lccc}
\hline
 &$WW^+ W^- \rightarrow e^\pm\nu e^\pm\nu$
&$H\rightarrow W^+ W^- \rightarrow e^+\nu e^-\nu$ 
&$H\rightarrow W^+ W^- \rightarrow e^\pm\nu \mu^\mp\nu$ \\
 &$WW^+ W^- \rightarrow  e^\pm\nu \mu^\pm\nu$
&$H\rightarrow W^+ W^- \rightarrow  e^\pm\nu \mu^\mp\nu$ 
& \\ 
 &$WW^+ W^- \rightarrow \mu^\pm\nu \mu^\pm\nu$
&$H\rightarrow W^+ W^- \rightarrow \mu^+\nu \mu^-\nu$ 
& \\  
\hline 
Luminosity (\ifb)         & 1.1 &  1.0 & 0.6\\ 
$m_{H}$ range (GeV/c$^2$)       & 120-200 & 120-200 & 120-200\\
Reference       & \cite{dzWWW} & \cite{dzHWWee,dzHWWmm} & \cite{dzHWWem-2b} \\
\end{tabular}
\end{ruledtabular}
\end{table}